\documentclass[global,twocolumn]{svjour}

\usepackage{graphicx}
\usepackage{amsmath}

\journalname{Applied Physics B}

\begin{document}

\title{Extended temperature tuning of an external cavity diode laser}

\author{C. S. Fletcher\inst{1} \and J. D. Close\inst{1}}

\institute{Department of Physics, Faculty of Science, The Australian National University}

\date{Received: date / Revised version: date}

\maketitle

\begin{abstract}
We describe the construction and operation of an external cavity diode laser in which the diode is cooled to $- 45^{\circ} \textup{C}$. This technique allows us to pull the wavelength of a nominally 782 nm diode to operate at 766.7 nm, a change of over 15 nm. The important aspects of our design are its ease and low cost as compared to other designs of changing the operating wavelength of laser diodes. We provide background information on thermal design with multiple TECs so the reader can extend this example for use in their own applications.
\end{abstract}

\section{Introduction}
\label{Introduction}

% Key point: laser diodes are cheap, easy to use. Want to make a system that maintains these qualities while allowing us to change the wavelength of the laser diode %

% Diode lasers are cheap, readily available and used widely in physics %
Diode lasers are ubiquitous in many areas of physics and engineering. Their production is driven by their use in the telecommunications industry, but their resulting low cost and availability, and their ease-of-use, has seen their implementation in many areas of physics, such as atom optics and high resolution spectroscopy \cite{WiemanLasers}.

% Diode lasers don't come at all wavelengths  - but sometimes we need them %
Although diode lasers are cheap and readily available at many wavelengths across the visible and near infrared part of the spectrum, there are many intermediate wavelengths that are not produced commercially. Many of these intermediate regimes, although of no interest to big consumers such as the telecommunications industry, are of vital interest to researchers working in atom optics and spectroscopy.

% We have ways and means! %
Since diode lasers were developed, techniques have been developed to change their operating characteristics from their nominal values to something that better suits a particular user. Their low cost, ease of use and reliability often makes performing such engineering feats preferable to pursuing other technologies. However, as the modifications to a diode laser system become more complicated we risk losing these benefits.

% We use ECDLs.%
In atom optics one of the best examples of well engineered systems based around mass-produced laser diodes is the external cavity diode laser (ECDL) \cite{WiemanECDL,HanschECDL}. It is capable of narrowing the spectral output of typical ``single mode" laser diode to less than 1 MHz, while maintaining a tuning range of up to $\pm$ 2 nm.

% Wavelength pulling etc. %
For experiments at wavelengths 10 nm or more from a commercially produced diode much more equipment, expertise or expense is required to harness these useful tools. At some significant cost diodes can be custom made or they can be wavelength selected from the very small number  of commercially produced diodes that lase more than 10 nm from their nominal design wavelength. Mass produced diodes can be anti-reflection coated in small batches and built in to an external cavity configuration to force them to lase $\pm$ 5 nm from their free-running wavelength \cite{WiemanLasers}.

% Temperature tuning - general %
Alternatively, the temperature of the diodes can be reduced, changing the position of the gain curve and microscopically shrinking the laser cavity, reducing the operating wavelength of the diode. Typical diode lasers have a temperature tuning coefficient of 0.25 nm/K \cite{WiemanLasers}. In many experiments, the diode is temperature controlled using a thermoelectric cooler in a feedback configuration to prevent unwanted temperature tuning of the diode. This technique is often coupled with some small amount of constant cooling, or heating, to ``pull" the diode wavelength several nanometers around its design wavelength.

% Temperature tuning - other techniques %
To pull the wavelength of a typical laser diode more than 15 nm, its temperature must be reduced approximately $60^{\circ} \textup{C}$. This sort of cooling has been achieved in the past by combinations of closed-cycle refrigeration, liquid nitrogen cooling, or multistage thermoelectric cooling and vacuum or nitrogen purged hermetically sealed chambers \cite{WeidmannLN2,CarterCCD}. These techniques are also used in other fields, such as cooled CCD arrays and cryogenic photodetectors and amplifier electronics. These techniques are relatively complicated, expensive and potentially unreliable, defeating many of the benefits of using cheap, rugged laser diodes.

% Temperature tuning - our technique %
This paper explores an extremely economical, rugged and simple design for cooling a diode laser sufficiently to pull its wavelength 15 nm below its room temperature design wavelength using only two thermoelectric coolers and judicious but simple insulation techniques. It is cheaper than the complicated systems used in other experiments, requires almost no maintenance, and performs sufficiently well to be used in real atom optics experiments.

\section{The Anatomy of an Ultracold System}
\label{lasersystem}

% Intro - key parts of the ultracold laser

\begin{figure}
\begin{center}
\resizebox{0.4\textwidth}{!}{\includegraphics{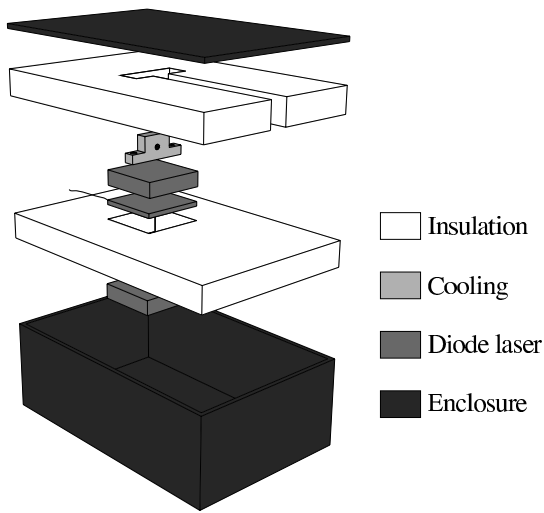}}
\end{center}
\caption{An ultracold diode laser system. The key components of an ultracold system are the insulation, the cooling and the sealed laser enclosure.}
\label{Fig_simplelaser}
\end{figure}

Figure \ref{Fig_simplelaser} shows an ``ultracold" diode laser system, although the basic components are the same in all efficient cooling devices: the cooling system, the insulation and the sealed enclosure. Each of these components has been implemented in various ways in other ultracold devices. However, we show that by judicious design and understanding of the system, we can construct a simple, cheap ultracold diode laser using solid-state components requiring low maintenance. We outline below our methods of implementing each component of the ultracold system. We go on to provide quantitative background on each of these components that will allow the reader to extend our design to more general systems, then show how we applied the method successfully to produce an ultracold external cavity diode laser that operates more than 15 nm below its room temperature wavelength.

	\subsection{Cooling System}

% How and why we used TECs

Thermoelectric coolers are heat pumps that move heat from their cold face to their hot face. If the hot face is attached to a heatsink held near room temperature, the cold face can fall up to $70^{\circ} \textup{K}$ below the ambient temperature, although more typical experimental values are of the order of $40^{\circ} \textup{K}$. In order to reduce the temperature of a laser diode by more than $60 ^{\circ} \textup{K}$ as well as removing heat generated in the diode, it is necessary to use multiple stages of TEC cooling.

In many ways TECs are the ideal device for cooling and temperature stabilizing a laser diode. They are small, cheap, cause no vibration and are reliable and maintenance free. Unfortunately they are also very inefficient, creating many times as much heat as they can transport away from their cooled surface. Removing this large amount of heat from the system requires very careful design of the entire TEC system, including selection of specific TECs, design and use of ``cold plates" to allow sufficient insulation between stages, and selection of the heatsink. These factors are modeled roughly and considered further in Section \ref{TECs}.

	\subsection{Insulation}
	
% How and why we used polystyrene

Any unnecessary coupling of heat in the system is bad, but unnecessary coupling at the coldest stage is the most wasteful because it must then be pumped out of the system, and magnified many times, by the inefficient TECs. The laser diodes we use in our laser cooling and trapping experiments typically dissipate less than 200 mW as an active heat load, but the passive heat flow from the surrounding laboratory, due mostly to convection, can act as a much larger heat load. Insulation is necessary to reduce these passive heat loads to reasonable values. Some ultracold systems have been built in vacuum chambers to minimize passive convective heat transfer, but in Section \ref{Insulation}, we show that simple design and careful construction can yield a straight-forward polystyrene insulation scheme that is sufficient for our ultracold system.

	\subsection{Sealed Enclosure}

% How and why we used a die-cast enclosure at IP65

As an ultracold system is cooled below room temperature, condensation will form on the coldest parts of the system, which as the temperature drops becomes ice. In more complicated ultracold systems this too has been overcome by enclosing the laser in a vacuum chamber. However, in Section \ref{Enclosure} we show that for this sort of laser, a simple IP65 neoprene-sealed cast aluminum jiffy box is a far cheaper and simpler solution.

	\section{Thermoelectric coolers}
	\label{TECs}

%		Perfect for laser cooling 'cause solid state, no vibration etc.
%		Very inefficient, so need to design stack carefully
%		"Design" implies choosing the correct TECs from a list from manufacturer
%		Choosing TECs implies choosing n and G
%		Man. data sheets yield most efficient TECs ... not necess. what we want
%		We want to optimize Tmin, can make model

\begin{figure}
\begin{center}
\resizebox{0.4\textwidth}{!}{\includegraphics{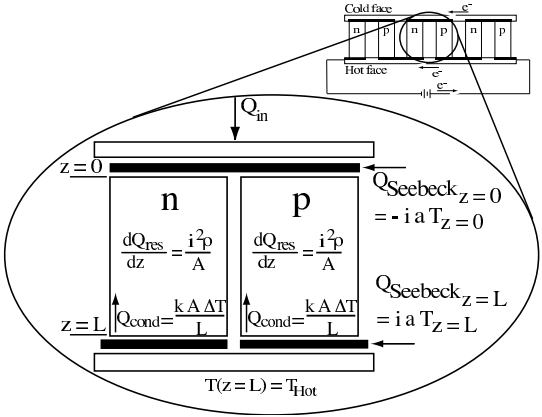}}
\end{center}
\caption{The pertinent heat flows in a single junction of a TEC: resistive, conductive and thermoelectric. The thermal equations noted below apply in the middle region of the TEC - the localized Seebeck heating at each face is applied purely as a boundary condition to the standard equations. }
\label{Fig_heatmodel}
\end{figure}

% What we are doing - deriving our own eqns
% Manufacturer's web site
% References

TECs are very inefficient, and the design of a multiple TEC stack is critical in ensuring any sort of useful operation of a given system. Unfortunately, manufacturer's handbooks and web sites seem biased towards engineering issues such as efficiency of single TECs, or specific extrema of performance, such as maximum temperature differential with zero heat load \cite{melcor,ferrotec}. Published works on multi-TEC stacks are understandably general in nature, and often interested in engineering figures of merit such as cost of construction and running cost. By generating our own simple model, we can perform ``custom" optimizations on specific systems such as a multi-stage TEC stack for ultracold ECDLs, or even just choosing a TEC under non-standard limiting conditions, e.g. for use with a specific temperature controller.

% Key choices - n and G ... %

Very little information seems to be available on modeling and choosing TECs for a given application, especially for multi-TEC stack configurations. All TEC manufacturers use semiconductors with similar performance at room temperature. Any given TEC has well-defined optimum voltage and current for any given configuration. The only options available to the user buying TECs off-the-shelf are the ``shape factor", $\textup{G = Area / Length}$ of individual thermocouples (0.01 - 1 cm), and n, the number of thermocouples in the TEC (20 - 200). However, even with stacks of only two TECs, the optimization of n and G for each TEC becomes non intuitive and must be modeled.  Selection criteria for these parameters are derived below.

	\subsection{A Thermal Model}

Figure \ref{Fig_heatmodel} shows a thermal model for a single thermocouple system. The thermoelectric effect is quantified by the Seebeck coefficients of the semiconductors, $a$, which for commercially used materials is of the order of $2 \times 10^{-4} \textup{V}\textup{K}^{-1}$. Other heating effects inherent in the system are resistive heating inside the semiconductors, quantified by the resistivity, $\rho$, and heat flow through the semiconductors due to the temperature gradient created, quantified by the conductivity, $\kappa$. Equation \ref{heateqn setup} shows the 1D thermal model with boundary conditions.

\begin{equation}
\begin{split}
& - \kappa G L \frac{\partial^{2} T}{\partial z^{2}}  = \frac{i^{2} \rho}{G L} \\
B.C. 1:	& -\kappa G L \frac{\partial T}{\partial z} | _{z = 0}  = Q |_{z = 0} =Q_{in} + i a T_{c} \\
B.C. 2:	& \hspace{4mm} T |_{z = L}  = T_{H} \\
\end{split}
\label{heateqn setup}
\end{equation}

It is straight forward to solve this model and find the temperature at the cold side of the TEC in terms of the TEC parameters, the heat load on the system, $Q_{in}$, and the hot side temperature, $T_{H}$. Note that Equation \ref{heateqn temp} extends the model above to $2 n$ junctions, as commercial TECs are typically constructed from 20 - 200 semiconductor junctions.

\begin{equation}
T_{C} = T |_{z = 0} = \frac{T_{H} + \frac{i^{2} \rho}{2 \kappa G^{2}} + \frac{Q_{in}}{2 n \kappa G}}{1 + \frac{i a}{\kappa G}}
\label{heateqn temp}
\end{equation}

	\subsection{Deriving Standard Equations}

Very quickly, we can make some general statements about the design of TEC-based systems by investigating Equation \ref{heateqn temp} in various limiting cases. The heat that can be pumped from a system attached to the TEC cold face is $Q_{in}$. The heat that a TEC expels from its hot face is $Q_{out}$. $T_{C}$ for a single TEC attached to a perfect heatsink is minimized (assuming our TEC has fixed $T_{H}$) with a current of $i_{opt}$.

\begin{equation}
\begin{split}
Q_{in} & = 2 n ( i a T_{C}  - \frac{i^{2} \rho}{2 G} - \kappa G \Delta T ) \\
Q_{out} & = Q_{in} + \frac{2 n i^{2} \rho}{G} + 2 n i a \Delta T \\
i_{opt} & = \frac{G \kappa}{a} (\sqrt{1 + \frac{2 a^{2}}{\rho \kappa} T_{H}} - 1) \\
\end{split}
\label{heateqn derived}
\end{equation}

These equations contain information useful in developing an intuitive understanding of how systems using TECs should be designed. $Q_{in}$ is maximized when $\Delta T = 0$. $Q_{out}$ is simply $Q_{in}$ plus the resistive heating inside the semiconductor legs of the TEC, plus the difference between the heat absorbed by the Seebeck effect at the cold face and the {\em heat liberated at the hot face}. The heat liberated at the hot face does not enter Equation \ref{heateqn temp} above because we are assuming a perfect heatsink that can maintain $T_{H}$ independent of heat load, but the more advanced models in Section \ref{heatsink} do require consideration of this heat load. 

We are pushing the limits on the lowest temperatures that can be achieved using simple TEC stacks and polystyrene insulation. One measure of how well we can do this is the lowest temperature we may expect to achieve for a given system. Equation \ref{heateqn temp} yields this quantity directly for given choosable TEC paramters (n and G), given system parameters ($Q_{in}$ and $T_{min}$), and a given operating condition (i). Equations \ref{heateqn derived} yield other important information such as the optimum operating current and the amount of heat output from the system. 

\begin{figure*}
\begin{center}
\resizebox{0.9\textwidth}{!}{\includegraphics{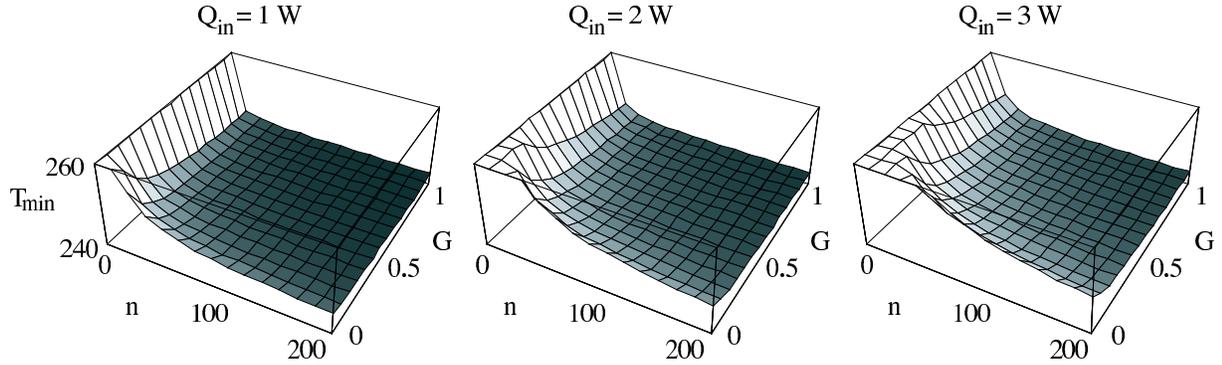}}
\end{center}
\caption{$T_{min}$ versus n and G for a series of system configurations with fixed $T_{Hot} = 300 \textup{K}$.}
\label{Fig_results1}
\end{figure*}

Figure \ref{Fig_results1} shows a series of plots of $T_{min}$ at $i_{opt}$ versus n and G, at various $Q_{in}$ typical of our system. The key result to note is that except for the very smallest TECs (a few, small thermocouples = low n, low G) most selections of TEC will yield the theoretical best $T_{min}$, approximately $70^{\circ}\textup{K}$ below room temperature. This tells us that our expected loads of $\sim 1 W$ of heating (investigated more fully in Section \ref{thermalloads}) are low compared to the design specification of most commercially available TECs. So should we just choose the largest TEC available to ensure that we can maintain $\Delta T \sim 70^{\circ}\textup{K}$? No, there is more to the calculation, as shown below.

	\subsection{Extensions to the model I - Real heatsinks}
	\label{heatsink}

An important example of the necessity of deriving our own equations is the extension of equation \ref{heateqn temp} to include a real heatsink. As it stands, equation \ref{heateqn temp} assumes that $T_{H}$ is ``set" by attaching the TEC to a reservoir, in practice some sort of heat sink. Heatsinks come in many shapes and sizes. Typically one thinks of a finned block of black aluminium, which loses heat to its surrounding environment ``efficiently" due to predominantly convective losses from its large surface area. By efficient, we mean that the heatsink can be only slightly hotter than the surrounding environment while losing a lot of heat due to this temperature differential. This property can be quanitified as ``thermal resistance", $\Theta [^{\circ} \textup{K/W}]$, which yields a number describing how much hotter the heatsink must get to dissipate an extra watt of power. A reservoir, or ideal heatsink, has $\Theta = 0^{\circ} \textup{K/W}$ so that it can maintain ``room" temperature independant of the power it must dissipate. A normal, passive aluminium heatsink as described above has $\Theta \sim 1^{\circ} \textup{K/W}$, at best. A standard forced air aluminium heatsink can have as low as $\Theta \sim 0.3^{\circ} \textup{K/W}$ - more advanced designs can perform even better.

These considerations are actually vitally important, as a well-designed multi-stage TEC system rapidly ends up limited by the heatsinking arrangement implemented. If the heatsink temperature rises due to the heat being dissipated, it pulls $T_{H}$ up with it. Since a TEC at best maintains a temperature differential, $\Delta T$, across its faces, $T_{C}$ rises correspondingly. Hence, the thermal resistance of the heatsink directly determines how cold our system can get. Increasing the size and power of a TEC in a heatsink limited system increases $T_{min}$.

Including a heatsink in our TEC model is easy using a mathematical software package, such as Mathematica\texttrademark. Instead of using a fixed $T_{H}$, we make it dependent on $Q_{out}$, that is, $T_{H} = T_{room} + \Theta Q_{out}$ for a heatsink with thermal resistance $\Theta$.

\begin{figure*}
\begin{center}
\resizebox{0.9\textwidth}{!}{\includegraphics{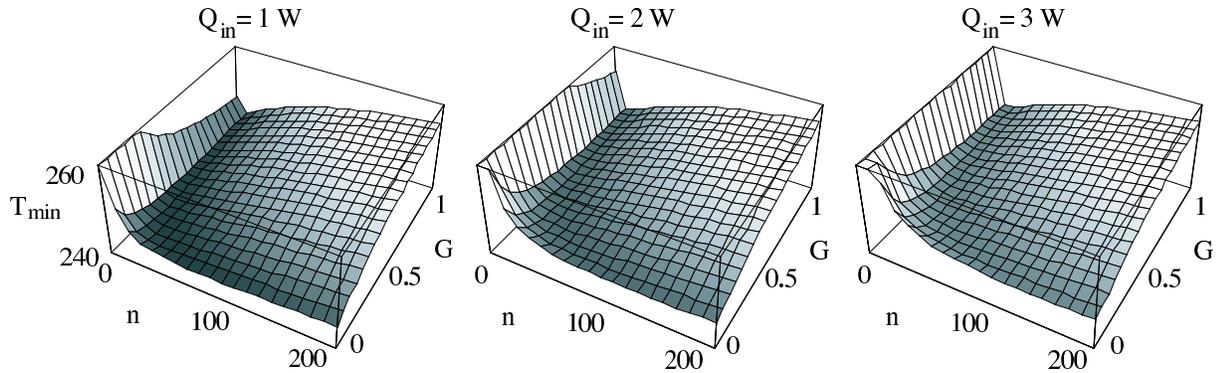}}
\end{center}
\caption{$T_{min}$ versus n and G for a series of system configurations using a real heatsink with thermal resistance $\Theta = 0.3 ^{\circ}\textup{K/W}$ with a room temperature of 300 K.}
\label{Fig_results2} 
\end{figure*}

Figure \ref{Fig_results2} shows plots similar to Figure \ref{Fig_results1}, but includes the effect of using a real heatsink. Obviously this effect is very important in selecting the TEC for our application. Even though Figure \ref{Fig_results1} indicated that any ``large" TEC would reach the theoretical lowest possible temperature in our system, Figure \ref{Fig_results2} shows that the extra heat load generated by a large TEC ends up raising the temperature of the heatsink so much that it degrades the final performance of the system as a whole. There is a clear optimum TEC for the system, albeit quite weakly dependent on the shape factor, G, and number of the thermocouples, n. This relative insensitivity to n can allow us to match the impedance of a TEC (drive voltage/current) to available TEC power supplies.

	\subsection{Extensions to the model II - Multi-stage systems}
	\label{multi}

\begin{figure}
\begin{center}
\resizebox{0.4\textwidth}{!}{\includegraphics{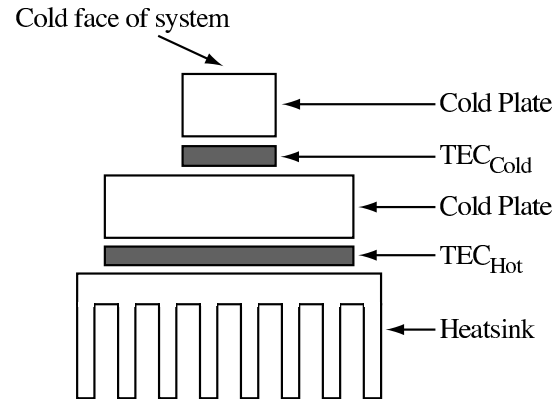}}
\end{center}
\caption{Multiple TECs stacked in series to provide total temperature drops of greater than $70^{\circ}\textup{K}$. Note the use of cold plates to allow extra thickness of insulation between TEC stages (insulation not shown). }
\label{Fig_multitec}
\end{figure}

% Multistage TECs %
Figure \ref{Fig_multitec} shows how TECs can be run in series in a ``multistage" configuration to provide greater temperature drops than a single unit. Their inefficiency puts severe limitations on how many stages should be used, because each extra stage must remove both the heat pumped, and more significantly, the heat generated by the previous TEC. Of course, the final stage of the system, the heatsink, must dissipate all this heat efficiently to yield the lowest possible temperature at the cold point of our system. Any unnecessary inefficiency in our choice of the number of stages, or the type of TEC to use in each stage of our system will directly impact the ultimate temperature we can achieve.

The theory presented above can be extended to multiple stages of TEC cooling. Equation \ref{heateqn temp} is set up for each stage. Assuming the TECs are connected by negligible thermal resistance, and there is suitable insulation from external heat loads, all the heat from the output of a given TEC stage must be transmitted to the input of the next stage. Similarly, the output of a given TEC stage should be at the same temperature as the input of the next TEC stage. The equations are straight-forward but can be quite detailed and are best dealt with in a mathematical package.

Optimization of a multi-stage system is somewhat difficult due to the size of the parameter space available to probe. However, in many realistic experimental situations not all parameters will be truly free; you may need to operate with a given TEC power supply that will limit both the drive current available and the number of thermocouples in a given TEC. In practice, using some combination of the general design techniques for individual TECs elucidated above, integrating the limitations imposed by your specific experiment, and then optimizing over the remaining available parameters computationally yields a workable technique for selecting TECs in multi-stage configurations.

	\section{Insulation}
	\label{Insulation}
	
% Operating at limits of system and heat load directly increases T_cold => need to minimize heat at input
Section \ref{TECs} showed that a compromise existed between the ultimate cold face temperature and the heat pumped through the cold face of a given TEC, that TECs are very inefficient, producing much more heat than they pump, and that most well-designed multi-stage TEC systems were limited by the heat dissipated at the heatsink of the system. Clearly, the lowest ultimate operating temperature of a TEC system will be very strongly affected by the amount of heat being pumped at the coldest face in the system.
	
	\subsection{Thermal Loads}
	\label{thermalloads}
% Active versus passive %
The laser diodes we use in our laser cooling and trapping experiments typically dissipate less than 200 mW as an active heat load, which makes them ideal for use in a thermoelectric cooler system. However, the passive heat flow from the surrounding laboratory to the significantly colder laser diode, predominantly due to convection heating via air, can act as a much larger heat load. Preventing heat transfer by convection is part of the reason ultracold systems are sometimes built in vacuum chambers.

% Passive head load estimates %
The passive heat loads in a real system can be estimated quantitatively from the very simple formulae in Equation \ref{loadeqns} for heat flow due to radiation, conduction and convection \cite{melcor}.

\begin{equation}
\begin{split}
Q_{conv} & = h A \Delta T \\
Q_{cond} & = \frac{\kappa A \Delta T}{\Delta z} \\
Q_{rad} & = \epsilon \sigma A (T_{hot}^{4}-T_{cold}^{4}) \\
\end{split}
\label{loadeqns}
\end{equation}

Typical values for the constants are $h=10 \textup{W}\textup{m}^{-2}\textup{K}^{-1}$ (rough value for still air), $\kappa=386 \textup{W}\textup{m}^{-1}\textup{K}^{-1}$ (copper), $\epsilon = 1$ (worst case), $\sigma=5.67\times10^{-8} \textup{W}\textup{m}^{-2}$. The value of A is the exposed surface area of the cold device, $\Delta z$ is the thickness / length of the conducting medium.

For a 2 cm x 2 cm x 2 cm copper block exposed on five faces, cooled 60 C below a room temperature of 20 C, connected to six 30 AWG copper wires at room temperature, the loads are approximately 1.2 W, 0.001 W, and 0.5 W respectively. Obviously, the convective and radiative loads are a severe limitation on the operation of the thermoelectric cooler system.

	\subsection{Insulation Design}
	\label{InsulationDesign}

\begin{figure}
\begin{center}
\resizebox{0.4\textwidth}{!}{\includegraphics{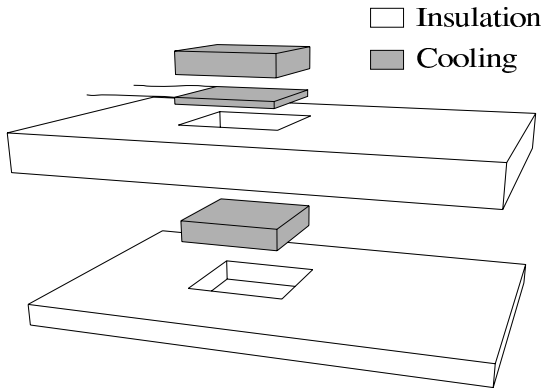}}
\end{center}
\caption{The use of a cold plate to increase the distance between the hot and cold surfaces either side of a TEC is a well known and vital part of TEC system design. In an ultracold ECDL the straightforward implentation of this simple idea is critical to the optimum performance of the cooling system.}
\label{Fig_coldplate}
\end{figure}

% Insulation %
These loads can be greatly reduced by the judicious use of insulation around the cooled components. Although this concept is straightforward, its importance cannot be overstated. Proper design and implementation of the the insulation around an ultracold laser is critical, and small changes can change the ultimate minimum temperature of the system by more than 20 K. However, it is vital to note that by a combination of following some general design principles calculated below and using careful but simple construction techniques it is possible to quickly, ruggedly, and extremely economically transform a standard ECDL into an ultracold laser lasing 15 nm below its design wavelength.

% Insulation estimates  %
We can estimate the amount of insulation required using the formulae above. By cladding our ultracold diode laser in polystyrene insulation (white, closed cell foam, conductivity $0.04 \textup{W}\textup{m}^{-2}\textup{K}^{-1}$) the convective and radiative loads calculated above are essentially removed, replaced by a conductive heat load through the insulation. We can calculate the amount of heat passing through a slab of polystyrene of area $10 \textup{cm}^{2}$, $t \textup{cm}$ thick due to a heat gradient of $60^{\circ}\textup{K}$, to be approximately $0.5/t \textup{W}$. This shows that simply cladding the coldest part of the diode laser in a couple of centimeters of polystyrene insulation will reduce the passive loads on the system to the order of magnitude of the active load. More accurate modeling of the heat flows in the system is simply not required, due to the experimental realities of constructing such a system. However, it is important to note the effect of several typical difficulties faced when constructing such a system.

% Insulation between TEC sides %
Typical TECs are approximately 3 mm thick, and can hold a temperature differential of up to 40 K in a realistic experimental setup. If $10 \textup{cm}^{2}$ hot and cold surfaces attached to such a TEC are insulated with 3 mm of polystyrene, the conduction between the surfaces can be calculated from the equations above to be approximately 0.5 W. Clearly this is another significant load on the cold face of the system, but again, careful design of the insulation is sufficient to minimize the problem. "Cold plates", such as those shown in Figure \ref{Fig_coldplate}, are copper or aluminum spacers, the same size as the face of the TEC, that allow more insulation to be added between the cold and hot faces of the system. Two 1 cm cold plates in the above example, allowing 23 mm of insulation, reduces the load on the system to less than 70 mW, which is much less than the active load of the diode itself. 

% Holes and wires %
As desirable as it might be, it is usually impossible to encase an ultracold system completely in insulation. An ultracold diode laser needs electrical wires attached to the diode itself, and a hole from the diode through the insulation to let light out of the system. Electrical wires for passing low-power signals into the system are normally not problematic - by using very fine gauge wires the conduction heat load in to the system can be kept small (e.g. 0.01 W for 6 30 AWG wires as calculated above), and even very fine gauge wires have low resistive heating - 10 cm of 30 AWG copper wire has a resistance of the order of $0.05 \Omega$, so could pass an amp of current with negligible heating. Holes in the insulation connecting the cold face of the system with the outside world are a serious problem, however. In practice this can end up a significant limitation on the performance of the cooling system, so holes should be kept to an absolute minimum, both in size and number.

	\section{Sealed Enclosure}
	\label{Enclosure}

As an ultracold system is cooled below room temperature, condensation will form on the coldest parts of the system which, as the temperature drops, becomes ice. Although this rarely permanently damages a hermetically sealed laser diode package, it can impede the operation of an external cavity laser diode system if there is a large build up of ice that scatters light from the main beam path.

Water will tend to condense on the first part of the laser enclosure that drops below the dew point of the system as the laser is cooled. In our case this is not the laser diode itself, but the copper block in which the diode laser is mounted, and any exposed parts of the peltier. However, on an average $20^{\circ}\textup{C}$ day with a relative humidity of 50 percent, the room air will contain about $8.7\textup{g}\textup{m}^{-3}$ of water. If we seal a 10 cm by 10 cm by 10 cm enclosure under these conditions, it will contain approximately $10^{-2}\textup{g}$ of water. As ice, this will occupy a volume of less than $5\textup{mm}^{3}$. By ensuring that the first point in the enclosure to pass the dew point is a relatively large area away from the direct path of the laser light - in our case the 10 mm x 10 mm top of laser diode block, we can guarantee that condensation will not significantly effect the operation of our laser.

This fact effects the design of our laser in one important way. The enclosure that we use around the laser must be relatively air-tight. The pressure of water vapor inside the enclosure drops when it condenses as the laser is cooled, and unless the enclosure is sealed water vapor will be pumped from the surrounding room air into the chamber. This will cause a large amount of ice to form inside the laser system, eventually impeding its operation. In practice, we have found a simple die-cast aluminum jiffy box sealed to IP65 gives many months of hassle-free operation.

\section{A Solid-State, Economical Ultra-Cold Diode Laser}
\label{Our Laser}

\begin{figure}
\begin{center}
\resizebox{0.4\textwidth}{!}{\includegraphics{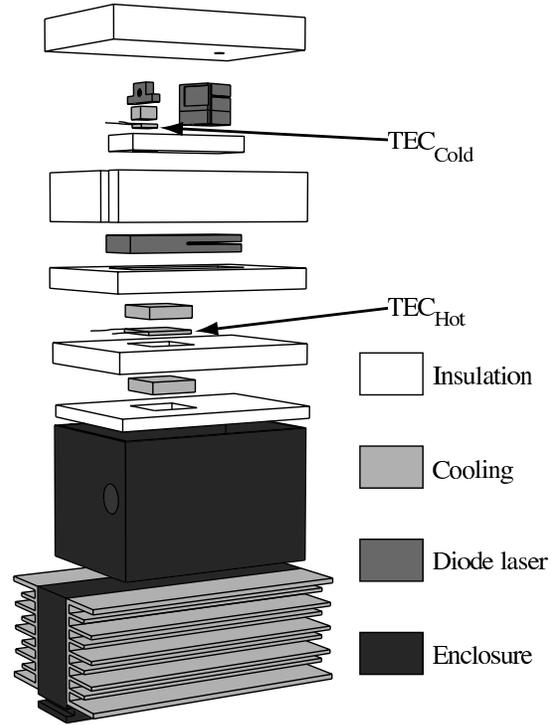}}
\end{center}
\caption{The ultracold external cavity diode laser system. The key components of an ultracold diode laser system are the insulation, the multi-stage thermoelectric cooling and the hermetically sealed laser enclosure.}
\label{Fig_ourlaser}
\end{figure}

% Laser overview %
Figure \ref{Fig_ourlaser} shows a diagram of the ultracold external cavity diode laser system that we have constructed using the principles outlined throughout this paper. It uses a \$ 50 Sanyo DL-7140 laser diode (75 mW) wavelength selected at 782 nm in a standard ECDL configuration \cite{HanschECDL}. The cooling system is comprised of two thermoelectric coolers (TECs) in a multi-stage configuration, utilizing cold plates and a pair of fan-forced heatsinks to provide robust and quiet cooling of the ECDL. The insulation is constructed from 6 pieces of polystyrene, carefully cut to provide snug-fitting insulation between all components and the enclosure. The enclosure is a simple die-cast aluminum ``jiffy" box, sealed with a neoprene gasket to IP65.

\begin{figure}
\begin{center}
\resizebox{0.4\textwidth}{!}{\includegraphics{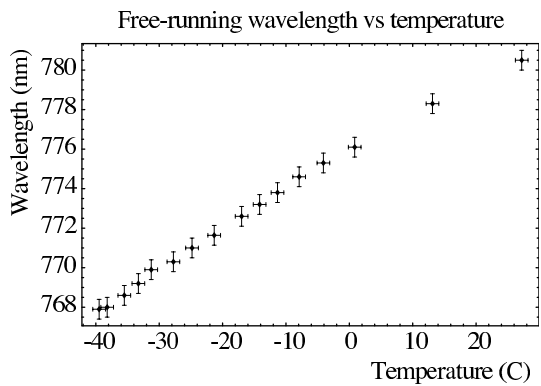}}
\end{center}
\caption{The temperature dependance of the free-running wavelength produced by a nominally 782 nm diode (at room temperature), as it is cooled to $-40^{\circ}\textup{C}$.}
\label{Fig_tempdepend}
\end{figure}

Figure \ref{Fig_tempdepend} shows the dependance of the diode laser operating wavelength on temperature. The wavelength was measured using a fiber-coupled ANDO AQ-6315A optical spectrum analyzer. The resistance of a thermistor mounted in the copper diode cooling block was measured and converted to a temperature using the manufacturer's specifications. Note that the figure shows data collected from a free-running laser diode, as opposed to one in an external cavity configuration. Figure \ref{Fig_sas} shows a saturated absorption spectrum \cite{WiemanECDL} of the D2 transition of potassium at 766.7 nm, collected using a cooled diode laser in an external cavity configuration. The data was collected on an uncalibrated photodiode as the laser wavelength was scanned by tuning the voltage across a piezoelectric transducer attached to the grating of the external cavity. This data shows stable, narrow linewidth operation of a diode laser at 766.7 nm, the wavelength we require for cooling and trapping experiments on potassium.

The key design specification of the cooling system was the ability to cool to $-40^{\circ}\textup{C}$ from a room temperature of $20^{\circ}\textup{C}$. There were several limitations on the design peculiar to our experiment. We needed to use a small capacity temperature controller built in to our laser controller for the cold TEC. We needed to be able to adjust two hex screws in the external cavity diode laser mount for alignment purposes, and we obviously needed to get the laser light out of the ultracold system. Also, as we were retro-fitting an existing ECDL, the physical size of the TECs had to be consistent with the rest of the laser. The design was based around two TECs, a small ``cold" TEC that cooled just the laser diode block, and a much larger ``hot" TEC that cooled the ECDL base plate.

\subsection{Cooling}

\begin{figure}
\begin{center}
\resizebox{0.4\textwidth}{!}{\includegraphics{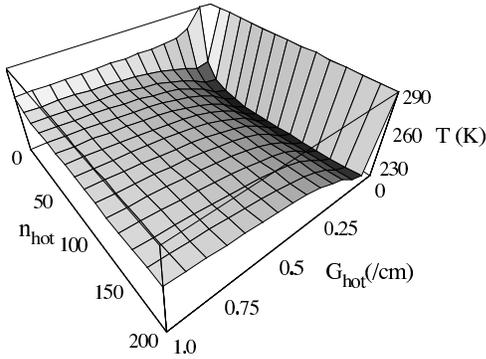}}
\end{center}
\caption{Minimum temperature versus $n_{Hot}$ and $G_{Hot}$ for our real ultracold diode laser system, including all its limitations. There is a clear minimum at $G \sim 0.1$.}
\label{Fig_ourresults}
\end{figure}

A general model for a two stage TEC system attached to a real heatsink was constructed as outlined in Section \ref{multi}, yielding an equation for $T_{C}$ in terms of the system parameters ($Q_{in}$ and $\Theta_{heatsink}$), the choosable TEC parameters ($n_{Hot}$, $n_{Cold}$, $G_{Hot}$ and $G_{Cold}$), and the operating points ($i_{Hot}$ and $i_{Cold}$). We estimated the heat load on the cold face of the system (with the insulation outlined below) at somewhere around 0.5 - 1 W and the thermal resistance of the heatsinks we implemented was specified as approximately $0.4 \textup{K/W}$. The temperature controller available to us supplied 2 A at 2 V, so this severely limited our choice of the cold TEC and its operating point. In addition, the power supply for driving the hot TEC was relatively modest, supplying 4 A at up to 20 V. The model that was left to optimize involved the selection of the hot TEC ($n_{Hot}$ and $G_{Hot}$) within the accessible current ranges. This parameter space is quite easy to optimize, and Figure \ref{Fig_ourresults} shows a plot of the results of the model.

The TEC chosen was close to the optimum value of G (within the choices available), although the relative insensitivity to n allowed us to tune the drive impedance to our power supply. The final cooling system design obeys the general principles we would expect for a multi-stage TEC system, within the specific requirements of this experiment. The ``hotter" TEC ($n_{Hot} = 67$, $G_{Hot} = 0.118$, 40 mm square, running at 60 W) has significantly larger capacity to transport the heat generated at the laser diode as well as the heat generated internally by the ``colder" TEC ($n_{Cold} = 31$, $G_{Cold} = 0.08$, 15 mm square, running at 4 W). The upper TEC is controlled by the laser controller in a feedback loop governed by a thermistor mounted in the laser diode copper block , but the lower TEC free-runs at constant current. The heat sinks are the largest useful models we could find: $0.4^{\circ}\textup{C/W}$ ``tube" heatsinks cooled by forced convection provided by four 80 mm 12 V fans. They typically sit only 5 - 10 degrees above ambient room temperature. The cold face of the system can reach as low as $-45^{\circ}\textup{C}$.

\subsection{Insulation}

The construction of the insulation followed the straightfoward concepts outlined in Section \ref{Insulation}, but the specific implementation and quality of the assembly had dramatic effects on the performance of the ultracold ECDL. The enclosure was chosen to provide a reasonable amount (several centimeters) of insulation around the laser components. Multiple pieces of insulation were cut, and often re-cut, to provide a snug fit between the enclosure and various components.The consequent piecing together of the 3-dimensional foam-and-laser jigsaw puzzle was a ``once-only" event. Dismantling the laser inevitably required re-cutting some foam pieces to ensure a snug fit. However, careful construction paid off; a well constructed ultracold laser would reach temperatures up to $15^{\circ}\textup{K}$ below a poorly constructed specimen.  

% Hole for light %
An ultracold ECDL may produce light 15 nm below its design wavelength, but unless you can get that light out of the system it is of little use. Our ultracold ECDL has the smallest possible hole in the insulation that will still allow coarse wavelength tuning using the grating, approximately 8 mm in diameter and 50 mm long. Even slightly larger holes (12 mm diameter) significantly increased the ultimate minimum temperature at the laser diode (~5 K).

% Foam cutter %
The ability to quickly, easily, and most of all accurately and repeatably cut polystyrene significantly effected the quality and performance of the finished ultracold laser. We constructed a foam cutter board by passing a few amps of current through a 22 AWG nichrome wire perpendicular to a board similar to a draughtsman's table that enabled quick, square, accurate cuts that fit together tightly to produce a well insulated system.

% Hex keys %
ECDLs require some tweaking, typically through hex key driven adjustment screws mounted in the baseplate and grating mount (Figure \ref{Fig_ourlaser}). Normal steel hex keys would provide a large heat path into the cooled laser, so we developed an insulating hex key using two halves of a single key separated by a length of 8 mm diameter perspex rod. It passes through the die-cast aluminum enclosure via standard IP65 cable glands. These keys were simple to construct and insulated the laser so effectively that they can be left attached to the system permanently without affecting the cooling performance. In this configuration the end of the hex key attached to the system is also cooled. This is vital, as the thermal shock induced by touching a room temperature hex key to the laser induces so much frequency drift that it becomes impossible to tune the laser effectively.  

% Wires %
Electrical signals are passed in to the laser enclosure through standard multi-core cables inserted via a standard IP65 cable gland. Once inside the enclosure the wires are soldered to 30 AWG enameled copper wire that passes through the insulation to the cold face of the system. In addition, the cables that must be attached to the ultracold copper block that houses the diode can be lagged to the less cold ECDL baseplate to ensure that no unnecessary heat is transferred into the ultracold region.

\subsection{Enclosure}

The size of the enclosure was roughly selected based on readily available models and the rough size of the laser and insulation. Most electronics suppliers carry a line of enclosures fitted with neoprene gaskets to provide dust-and-moisture sealing to IP65. They also supply standard ``cable glands", plastic feed-throughs with a highly compressible o-ring that clamps down on the cable passing through it when tightened. Two of these cable-glands were used to pass electric wires in to the enclosure; another two were used to allow the perspex allen keys to pass in to the enclosure, and even be turned, without disrupting the sealing. The laser light was passed out of the system by gluing an anti-reflection coated window over a hole drilled in the enclosure and sealing it with silicone adhesive. Similarly, the thermoelectric coolers were attached to the heatsinks via a cold plate glued with silicone in to a precisely cut hole in the base of the enclosure.

\section{Conclusion}
\label{Conclusion}

\begin{figure}
\begin{center}
\resizebox{0.4\textwidth}{!}{\includegraphics{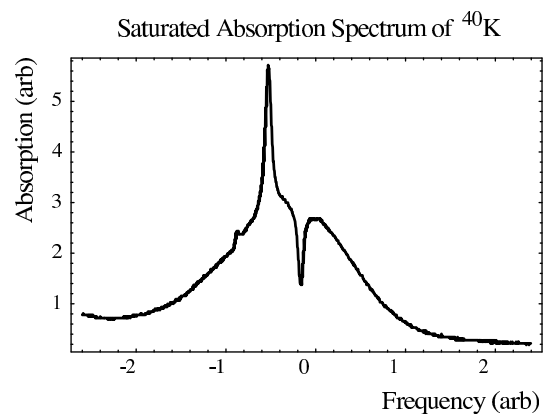}}
\end{center}
\caption{A saturated absorption spectrum of the potassium D2 line at 766.7 nm. The diode used to produce this spectrum ran at 782 nm at a room temperature of $20^{\circ}\textup{C}$, but has been cooled, in an external cavity configuration, to below $-40^{\circ}\textup{C}$ to reduce its wavelength.}
\label{Fig_sas}
\end{figure}

% What we have achieved %
We have demonstrated a simple, cheap and rugged design for pulling the wavelength of a laser diode in an external cavity diode laser 15 nm below its nominal design wavelength. Because our design only uses some carefully designed polystyrene insulation, a sealed die-cast aluminum box and one more TEC than a traditional ECDL, it maintains the inherent strength of the original ECDL design: a simple, rugged home-built structure that harnesses the cheap, easily integrated nature of mass produced telecommunications laser diodes to provide the sensitive optoelectronics of a laser. This makes it superior to complicated systems based on liquid nitrogen cooling or vacuum sealed cooled laser heads for many uses in atom optics and other physics laboratories. A nominally 782 nm laser diode has been cooled to below $-40^{\circ}\textup{C}$, and used to perform saturated absorption spectroscopy on the potassium D2 transition at 766.7 nm, as shown in Figure \ref{Fig_sas}.

In addition, we have presented background information on simple thermal design with thermoelectric coolers that would allow the interested reader to extend our example to other ultracold systems. Our method also allows the design of ultracold systems within project-specific limitations, such as working with legacy equipment, a common situation that many optimization-based references ignore.

\end{document}